\documentclass{article}
\usepackage{authblk}
\usepackage{geometry}
\usepackage{graphicx}
\usepackage{caption}
\usepackage{xcolor}
\usepackage{float}
\usepackage{soul}
\graphicspath{Images}
\usepackage{amsmath, amssymb}
\usepackage{hyperref}
\usepackage{cancel}
\usepackage[fleqn]{mathtools}

\usepackage{cancel}
\usepackage{chngcntr}
\counterwithin*{equation}{section}
\counterwithin*{equation}{subsection}
\renewcommand{\theequation}{\mbox{\arabic{section}.\arabic{equation}}}
\renewcommand{\theequation}{\mbox{\arabic{section}.\arabic{subsection}.\arabic{equation}}}
\usepackage [english]{babel}
\usepackage [autostyle, english = american]{csquotes}
\MakeOuterQuote{"}
\usepackage{relsize}
\usepackage{multirow}
\usepackage{bigints}
\usepackage{amsmath, amssymb, physics, bm}
\usepackage{hyperref}
\usepackage{amsthm}
\theoremstyle{remark}

\usepackage{tikz}
\usetikzlibrary{arrows.meta,calc}
\usepackage[toc,page]{appendix}

\usepackage{authblk}
\usepackage{orcidlink}
\usepackage{hyperref}
\usepackage{tabularx}
\usepackage{booktabs}
\usepackage{pdflscape}
\usepackage{empheq}

\title{Stationary Bohmian superposition under amplitude and phase modulation}

\author[1]{Anand Aruna Kumar\,\orcidlink{0000-0001-6148-2777}}

\affil[1]{Research Engineer, IBM Research, Albany, NY, USA\par
	\href{mailto:anand.aruna.kumar@ibm.com}{anand.aruna.kumar@ibm.com}}

\begin{document}
\date{} % Empty
\maketitle
\begin{abstract}
		\noindent In this work, we examine the problem of stationary superposition in the Bohmian amplitude--phase formulation, where amplitude and phase obey coupled nonlinear equations and direct linear superposition is not generally preserved. Considering two near-degenerate stationary branches, we derive a hierarchical reduction in which the mean amplitude satisfies an Ermakov--Pinney equation, while the difference amplitude evolves through a forced Mathieu--Hill type modulation induced by energy and stationary-current differences. It is shown that energy coherence alone does not uniquely determine phase coherence, since independent stationary currents continue to enter both the modulation and phase-difference equations. For weak amplitude modulation, a Wronskian-based stationary branch obtained from an Ermakov--Pinney solution admits a controlled amplitude--phase construction, leading to an algebraic phase representation and a Jacobi--Anger spectral expansion. As a result, a linear spectral structure emerges through Bessel-weighted amplitude and phase modulation. Such a representation is naturally suited for modelling aperture geometries, as illustrated by rectangular and parabolic slit reductions exhibiting Fresnel--type phase chirp and modulation-driven sidebands. The present construction therefore provides an analytical route by which linear spectral superposition re-emerges from nonlinear Bohmian amplitude--phase dynamics.
		\par	 
		\vspace{1pc}
     	\noindent \textbf{Keywords:}
     	 Bohmian mechanics; stationary superposition; amplitude and phase modulation; nonlinear superposition; Mathieu--Hill equation; Fourier--Bessel expansion
\end{abstract}

\section{Introduction}
\vspace{0.5pc}

\noindent
The Bohmian description of stationary quantum systems expresses the wavefunction in polar form
\(R\,e^{iS/\hbar}\), leading to coupled equations for the amplitude and phase
\cite{Bohm1952a,Holland1993}. Under stationary conditions, the continuity equation imposes a conserved current, while the Hamilton--Jacobi equation governs the phase evolution. Together, these reduce the amplitude dynamics to an Ermakov--Pinney structure \cite{Ermakov1880,Pinney1950}, whose associated nonlinear equation admits a first integral fixed by the Wronskian of two independent solutions of the corresponding linear Sturm--Liouville, or Helmholtz, problem \cite{SchuchD}. In this setting, the Wronskian \cite{Lewis1967,Guasti} acts as an invariant that constrains the nonlinear amplitude through the energy and continuity conditions.
\vspace{0.5pc}

\noindent
Unlike the linear Schr\"{o}dinger equation, the resulting amplitude--phase equations are nonlinearly coupled and do not generally preserve linear superposition at the level of the Bohmian variables. Although wavefunctions remain linearly superposable in the Hilbert-space representation, the corresponding amplitude and phase reorganise through nonlinear relations. The central question is therefore not whether superposition exists, but how linear spectral structure re-emerges from an underlying nonlinear stationary dynamics.

\vspace{0.5pc}

\noindent
This motivates the present formulation. We consider two near-degenerate stationary Bohmian branches with controlled amplitude and phase modulation. When the branches remain sufficiently close in energy and phase variation, their mean component preserves the stationary Ermakov--Pinney backbone, while the difference component evolves through a Mathieu--Hill type modulation. Conversely, when the amplitudes and phases become sufficiently dissimilar, the branches separate effectively and may be treated as independent stationary systems.

\vspace{0.5pc}

\begin{table}[h]
	\centering
	\caption{Outline of the stationary Bohmian superposition construction.}
	\label{tab:intro-approach}
	\begin{tabular}{p{0.26\linewidth} p{0.34\linewidth} p{0.30\linewidth}}
		\hline
		\textbf{Level} & \textbf{Structure} & \textbf{Role in the paper} \\
		\hline
		Stationary branch &
		Bohmian amplitude--phase decomposition with conserved current &
		Provides the Ermakov--Pinney backbone and Wronskian invariant \\
		
		Mean component &
		Near-degenerate averaged amplitude and phase &
		Preserves the stationary nonlinear structure \\
		
		Difference component &
		Small amplitude--phase deviation about the mean branch &
		Produces Mathieu--Hill type modulation \\
		
		Spectral representation &
		Fourier--Bessel expansion of the modulated phase factor &
		Restores linear superposition at the representation level \\
		
		Aperture geometry &
		Shifted one- and two-dimensional source constructions &
		Connects the formalism to slit-type interference and spatial ordering \\
		\hline
	\end{tabular}
\end{table}
\vspace{0.5pc}

\noindent
The question of superposition has also been explored in time-dependent Ermakov systems and trajectory-based formulations. In particular, Nassar and Miret-Art\'es \cite{NassarAB2017} examined superposition in Bohmian mechanics through trajectory analysis and open quantum systems, where coherence and interference arise through dynamical trajectory evolution rather than direct amplitude addition. The present work is complementary: it focuses on stationary near-degenerate branches and shows how spectral superposition can emerge from nonlinear amplitude--phase modulation. The overall reduction strategy adopted throughout the
manuscript is summarised in Table~\ref{tab:intro-approach}.

\vspace{0.5pc}

\noindent
The notion of controlled amplitude and phase modulation is not unique to foundational quantum descriptions. Related structures arise in optical systems, where beam shaping, coherent mode selection and interference control are achieved through phase and amplitude tuning using optical components and feedback electronics. Examples include mode locking, cavity stabilisation, phased optical arrays and controlled interference under aperture constraints. While the present construction is not intended as a direct model of such systems, it provides a mathematically analogous framework for describing stationary superposition under nonlinear constraints \cite{Haus2000,SalehTeich}.

\vspace{0.5pc}

\noindent
Subsequently, the superposition of two similar stationary systems is examined with emphasis on applications. In particular, a simplified two-slit analogue is developed via a parabolic slit reduction of a two-dimensional aperture geometry. In these examples, linearity emerges in the spectral representation, while the underlying dynamics retains a structured interplay between amplitude modulation and nonlinear phase evolution arising from the Bohmian energy and stationary continuity conditions. This description may be useful for analysing spatial coherence and ordering in light, matter waves and related macroscopic quantum systems.

\vspace{0.5pc}

\noindent
The paper is organised as follows. Section~\ref{sec:Sec2} develops the near-degenerate amplitude--phase decomposition and establishes the Ermakov--Pinney and Mathieu--Hill hierarchy. Section~\ref{sec:Sec3} derives the modulated branch and its Fourier--Bessel representation. Section~\ref{sec:Sec4} extends the construction to two-dimensional aperture geometries, while Section~\ref{sec:Sec5} develops the parabolic slit reduction and the spectral superposition of shifted one-dimensional sources. Section~\ref{sec:Sec6} provides a summary of the spectral organisation developed throughout the manuscript, followed by the conclusions in Section~\ref{sec:Sec7}. Technical details are collected in ~\ref{app:fourier-bessel}.

\section{Two near-degenerate stationary branches}
\label{sec:Sec2}
\vspace{0.5pc}

\noindent
We consider two stationary Bohmian branches in one dimension with nearly equal energies. Under a near-degenerate approximation, the amplitudes and phases admit a decomposition into mean and difference components, yielding a hierarchical structure in which the mean branch satisfies an Ermakov--Pinney equation. For constant potential and stationary current, the corresponding Ermakov solution for each branch may be written in Wronskian form as

\begin{equation}
	R^2(x)
	=
	A\sin^2(kx)
	+
	B\cos^2(kx)
	+
	2D\sin(kx)\cos(kx),
\end{equation}

\noindent where \(A,B,D\) are integration constants constrained
through the Wronskian invariant.
\vspace{0.5pc}

\noindent
The regime of interest lies between two limiting cases. For identical Wronskian amplitudes \((A=B)\), the stationary Ermakov solution reduces to a trivial constant-amplitude branch, while strongly dissimilar amplitudes or wave numbers \((A_1\not\sim A_2,\; k_1\not\sim k_2)\) lead to rapid phase separation and effective decoherence, making interference effects negligible. We therefore focus on weakly imbalanced stationary branches,
\[
B=A(1+\varepsilon),
\qquad
D=0,
\qquad
|\varepsilon|\ll1,
\qquad
k_1\simeq k_2.
\]
for which amplitude modulation remains tunable and superposition effects persist while retaining a controlled perturbative structure. The spectral reconstruction procedure for stationary
Bohmian branches is summarised in
Table~\ref{tab:sec2-reduction}.

\begin{table}[h]
	\centering
	\caption{Compact reduction scheme for near-degenerate stationary Bohmian branches.}
	\label{tab:sec2-reduction}
	\begin{tabular}{p{0.27\linewidth} p{0.62\linewidth}}
		\hline
		\textbf{Stage} & \textbf{Role in the reduction} \\
		\hline
		
		Near-degenerate branches &
		Introduce weakly imbalanced stationary branches with
		\(k_{1,2}=k_m\pm\Delta k/2\) and \(B=A(1+\varepsilon)\). \\
		
		Mean--difference split &
		Rewrite amplitudes and phases as mean and difference variables,
		separating the stationary backbone from the modulation branch. \\
		
		Mean branch &
		Adding the two Hamilton--Jacobi equations preserves the
		stationary Ermakov--Pinney structure. \\
		
		Difference branch &
		Subtracting the two equations gives the amplitude--phase
		modulation equation, with current and energy mismatch as sources. \\
		
		Weak Wronskian modulation &
		Using \(R_m^2=A(1+\varepsilon\sin^2(k_mx))\) reduces the
		difference branch to a Mathieu--Hill type equation. \\
		
		Spectral reconstruction &
		The modulated stationary branch is reconstructed spectrally in
		Section~\ref{sec:Sec3}. \\
		\hline
	\end{tabular}
\end{table}

\newpage
\subsection{Near-degenerate amplitude--phase decomposition}
Let the two stationary Bohmian branches be given by
\begin{equation}
\psi_{k_j}(x)
=
R_{k_j}(x)
\exp\!\left(
\frac{i}{\hbar}
S_{k_j}(x)
\right),
\qquad
k_1\simeq k_2.
\end{equation}

\noindent To describe weak modulation, we parameterise the near-degenerate
wave numbers as
\[
k_{1,2}
=
k_m
\pm
\frac{\Delta k}{2},
\qquad
|\Delta k|
\ll
k_m,
\]
so that the phase difference naturally encodes the detuning scale.
The two branches independently satisfy
\begin{equation}
	\frac{(S_{k_j}')^2}{2m}+V
	-\frac{\hbar^2}{2m}\frac{R_{k_j}''}{R_{k_j}}
	=E_j,
\end{equation}
and
\begin{equation}
	\label{eq:continuity1d}
	\frac{d}{dx}\left(R_{k_j}^2S_{k_j}'\right)=0.
\end{equation}
For the near-degenerate assumption 	$|\rho|\ll R_m$, $|\Delta S'|\ll |S_m'|$, the amplitudes and phases of each branch are given by
\begin{equation*}
	R_1=R_m+\rho,\qquad R_2=R_m-\rho,
\end{equation*}
and
\begin{equation*}
	S_1=S_m+\frac{\Delta S}{2},\qquad
	S_2=S_m-\frac{\Delta S}{2}.
\end{equation*}
\vspace{0.5pc}

\noindent The phase difference \(\Delta S\) captures the effect of a small wave-number detuning between the two stationary branches. For asymptotic free-particle phases one may write
\[
S_j'(x)\sim \hbar k_j,
\qquad
k_1\simeq k_2,
\]
so that
\[
S_m'
\simeq
\frac{\hbar}{2}(k_1+k_2),
\qquad
\Delta S'
\simeq
\hbar(k_1-k_2),
\]
prior to the additional amplitude--current coupling imposed by the Bohmian continuity relation.
\vspace{0.5pc}

\noindent Adding and subtracting the Hamilton--Jacobi equations, and retaining
leading-order terms under
\(
|\rho|\ll R_m
\)
and
\(
|\Delta S'|\ll |S_m'|
\),
gives the mean branch

\begin{equation}
\label{eq:meanbranch}
\boxed{
	\frac{S_m'^2}{2m}+V
	-\frac{\hbar^2}{2m}\frac{R_m''}{R_m}
	=E_m,
	\qquad
	E_m=\frac{E_1+E_2}{2}.
	}
\end{equation}
The difference  branch, neglecting $\rho^2$ term reduces to
\begin{equation}
	\label{eq:diffE}
		\frac{S_m'\Delta S'}{m}
		-\frac{\hbar^2}{m}
		\frac{\rho''}{R_m}+\frac{\hbar^2}{m}\frac{\rho R_m''}{R_m^2}
		=
		\Delta E.
\end{equation}
The continuity equation yields the phase-difference relation
\begin{equation}
	4R_m\rho\,S_m'
	+
	(R_m^2+\rho^2)\Delta S'
	=
	C_1-C_2.
\end{equation}

\begin{equation}
	\label{eq:neardegenphase2}
		\Delta S' \approx \frac{C_1-C_2}{R_m^2} - \frac{4C}{R_m^3}\,\rho.
\end{equation}
\noindent 
The modulation therefore separates into
current-induced and energy-induced contributions,
which remain coupled through the stationary continuity relation.
\vspace{0.5pc}

\noindent Substitution into the difference Hamilton--Jacobi  and Ermakov equations give

\begin{equation}
	\boxed{
		\rho''
	+
	\left[
	k_0^2+
	\frac{3C^2}{\hbar^2R_m^4}
	\right]\rho
	\simeq
	\frac{C(C_1-C_2)}{\hbar^2R_m^3}
	-
	\frac{m\Delta E}{\hbar^2}R_m .
	}
\end{equation}

\subsection{Small-amplitude Wronskian modulation}
For a constant potential \(V\) lower than the mean
energy of the two superposing stationary branches,
the corresponding mean wave number is
\[
k_m^2
=
\frac{2m(E_m-V)}{\hbar^2},
\]
with \(E_m\) denoting the average branch energy.
\vspace{0.5pc}

\noindent
To expose the modulation structure, we consider the Wronskian-based stationary branch generated by
\[
u_1(x)=\cos(k_mx),
\qquad
u_2(x)=\sin(k_mx),
\]
with Wronskian \(W=k_m\). The corresponding Ermakov amplitude takes the form
\[
R_m^2(x)
=
Au_1^2(x)
+
Bu_2^2(x),
\qquad
AB=\frac{C^2}{\hbar^2W^2},
\]
where the mixed term has been set to zero \((D=0)\).
\vspace{0.5pc}

\noindent
For weak amplitude modulation, we write
\[
R_m^2(x)
=
A
\left(
1+\varepsilon\sin^2(k_mx)
\right),
\qquad
|\varepsilon|\ll1,
\]
corresponding to \(B=A(1+\varepsilon)\). Using
\[
\frac{C^2}{\hbar^2A^2}
\simeq
k_m^2(1+\varepsilon),
\]
the amplitude-difference equation reduces, to first order in
\(\varepsilon\), to

\begin{equation}
	\label{eq:forcedmathieu}
	\rho''
	+
	\left[
	4k_m^2
	+
	3\varepsilon k_m^2
	\cos(2k_mx)
	\right]\rho
	\simeq
	F_C(x) - F_E(x),
\end{equation}
where the forcing induced by the mismatch of current and energy are respectively
\begin{equation}
	F_C(x)
	=
	\frac{C(C_1-C_2)}
	{\hbar^2A^{3/2}}
	\left[
	1-\frac{3\varepsilon}{4}
	+
	\frac{3\varepsilon}{4}
	\cos(2k_mx)
	\right] \text{ and} \quad F_E(x)= \frac{m\Delta E}{\hbar^2}\sqrt{A}\left[1+\frac{\varepsilon}{4}-\frac{\varepsilon}{4}cos(2k_mx)\right] .
\end{equation}
\noindent The corresponding phase-difference relation becomes
\begin{equation}
	\Delta S'
	\simeq
	\frac{C_1-C_2}
	{A\left(1+\varepsilon\sin^2(k_mx)\right)}
	-
	\frac{4C}
	{A^{3/2}
		\left(1+\varepsilon\sin^2(k_mx)\right)^{3/2}}
	\rho .
\end{equation}
\vspace{0.5pc}

\noindent To first order in \(\varepsilon\),
\begin{equation}
	\Delta S'
	\simeq
	\frac{C_1-C_2}{A}
	\left[
	1-\frac{\varepsilon}{2}
	+
	\frac{\varepsilon}{2}
	\cos(2k_mx)
	\right]
	-
	\frac{4C}{A^{3/2}}
	\left[
	1-\frac{3\varepsilon}{4}
	+
	\frac{3\varepsilon}{4}
	\cos(2k_mx)
	\right]
	\rho .
\end{equation}

\vspace{0.5pc}

\noindent In the energy-coherent sector \((E_1=E_2)\), the forcing associated with
energy mismatch vanishes, so that the modulation is driven entirely by
the stationary current mismatch \(C_1-C_2\).
\vspace{0.5pc}

\paragraph{Implications for stationary superposition.}

\noindent
\noindent
The preceding construction highlights an important distinction
between superposition in the Schr\"{o}dinger and Bohmian
representations. Although near-degenerate or energy-coherent
branches simplify the modulation dynamics, the nonlinear
coupling between amplitude and phase remains structurally
nontrivial through the stationary currents \(C_i\), which
continue to enter both the modulation equation and the
phase-difference relation. Consequently, superposition is not
naturally closed at the level of individual amplitude--phase
variables in the same sense as linear wavefunction addition in
the Schr\"{o}dinger picture. The resulting modulated amplitude $\rho$ acquires a Mathieu--Hill ~\cite{Hill1886,McLachlan1947,NayfehMook1979} type structure. However, even in the homogeneous limit, the nonlinear amplitude--phase coupling prevents a direct assessment of spectral linearity. We therefore develop in the following section an alternative spectral representation enabled by the weakly modulated stationary condition.

\section{Spectral representation of the stationary Bohmian branches}
\label{sec:Sec3}
The preceding analysis shows that direct amplitude--phase superposition of nearby stationary Bohmian branches does not naturally close, since phase evolution remains coupled to both amplitude modulation and stationary current mismatch. It is therefore advantageous to first construct an individual stationary branch with a consistent amplitude--phase structure and subsequently develop its spectral representation prior to superposition.
\vspace{0.5pc}

\noindent
For a stationary Bohmian branch,
\[
\psi_k(x)
=
R_k(x)
e^{\frac{i}{\hbar}S_k(x)},
\]
the stationary continuity relation gives
\begin{equation}
	S_k'(x)
	=
	\frac{C}{R_k^2(x)},
\end{equation}
so that the phase is determined nonlocally through
\begin{equation}
	S_k(x)
	=
	S_k(x_0)
	+
	\int_{x_0}^{x}
	\frac{C}{R_k^2(\xi)}
	\,d\xi.
\end{equation}

\noindent For the Wronskian-modulated stationary branch,
\begin{equation}
	R_k^2(x)
	=
	A_k
	\left(
	1+\varepsilon_k\sin^2(kx)
	\right),
	\qquad
	|\varepsilon|\ll1,
\end{equation}
the phase derivative becomes

\begin{equation}
	\label{eq:inv_rho2_eps_expand}
	S_k'(x)=	\frac{C}{R_k^2(x)}
	=\frac{C}{A_k}\,\frac{1}{1+\varepsilon_k \,sin^2(kx)}
	=\frac{C}{A_k}\Big[1-\varepsilon_k \, sin^2(kx)\Big].
\end{equation}
Integrating to first order in \(\varepsilon_k\) gives
\begin{equation}
	\label{eq:S_linear_leading}
	S_k(x)=S_{0k}+\hbar k x + \frac{\hbar\varepsilon_k}{4}sin(2kx),
\end{equation}
\noindent 
The stationary phase therefore consists of a carrier
contribution \(\hbar kx\) together with a weak
oscillatory modulation induced by the stationary
Wronskian imbalance.
\vspace{0.5pc}

\noindent	Using the small--order expansions
\begin{equation}
	R_k(x)\simeq
	\sqrt{A_k}\left[
	1+\frac{\varepsilon_k}{2}\sin^2(kx)
	\right],
	\qquad
	S_k(x)=\hbar kx+\frac{\hbar\varepsilon_k}{4}\sin(2kx),
\end{equation}
we obtain
\begin{equation}
	\psi_k(x)=
	\sqrt{A_k}
	\left[
	1+\frac{\varepsilon_k}{2}\sin^2(kx)
	\right]
	e^{ikx}
	\exp\!\left[
	i\frac{\varepsilon_k}{4}\sin(2kx)
	\right].
\end{equation}

\noindent	Using the Jacobi--Anger expansion~\cite{Arfken},
\begin{equation}
	e^{iz\sin\theta}
	=
	\sum_{n=-\infty}^{\infty}
	J_n(z)e^{in\theta},
\end{equation}
and, after rearrangement of trigonometric terms (see ~\ref{app:fourier-bessel}), we obtain
\begin{equation}
	\label{eq:jaexp1d}
	\psi_k(x)
	=
	\sqrt{A_k}
	\sum_{n=-\infty}^{\infty}
	\mathcal C_n(\varepsilon_k)\,
	e^{i(2n+1)kx},
\end{equation}
\noindent Here \(\mathcal C_n\) denotes the effective spectral
coefficient obtained as a linear combination of Bessel
contributions arising from the amplitude--phase
modulation.
\vspace{0.5pc}

\noindent An important advantage of the spectral representation is that the nonlinear phase modulation generates a discrete Fourier--Bessel structure with square-summable coefficients, defining momentum sidebands centred around the carrier wave number k. The corresponding Bessel coefficients satisfy
\[
\sum_{n=-\infty}^{\infty}
J_n^2(\varepsilon_k/4)=1,
\]
and the effective spectral coefficients obey
\[
\sum_n
|\mathcal C_n|^2
=
1+\varepsilon_k/2.
\]

\noindent Consequently, the real amplitude envelope fixes the spatial normalisation over an interval $L$ as
\[
A_k
=
\left[
L(1+\varepsilon_k/2)
\right]^{-1}.
\]
\noindent The nonlinear phase modulation therefore generates Bessel-weighted momentum sidebands centred around $\hbar k$, analogous to spectral broadening mechanisms encountered in nonlinear optical systems~\cite{Agrawal}.
\vspace{0.5pc}

\noindent The local intensity follows directly from the amplitude envelope,
\[
|\psi_k(x)|^2
=
R_k^2(x),
\]
yielding a smooth periodic modulation at twice the carrier frequency. The resulting spectral sidebands manifest primarily through momentum distributions and interference patterns arising from translated superpositions, rather than through direct modification of the local intensity envelope.

\section{Separable aperture geometries}
\label{sec:Sec4}
\vspace{0.5pc}

We extend the stationary Bohmian construction to two spatial dimensions,
focusing on rectangular aperture geometries. Throughout, we restrict
attention to the locally closed flux sector, ensuring compatibility with
standard quantum-mechanical observables while preserving the Ermakov
structure of the amplitude equations.

\subsection{Flux closure and Ermakov sector}
\vspace{0.5pc}
\noindent
For separable solutions, we impose the locally closed
flux condition, suppressing component-wise current
redistribution while preserving stationary directional
flux constants,
\[
R_i^2\,\partial_i S=C_{i0}.
\]
This reduces the continuity relation to a locally closed
sector compatible with standard quantum mechanics while
retaining the Ermakov--Pinney structure.
\vspace{0.5pc}

\subsection{Rectangular aperture geometry}
\vspace{0.5pc}

For Cartesian separability,
\begin{equation}
	R(x,y)=X(x)Y(y),
	\qquad
	S(x,y)=S_{k_x}(x)+S_{k_y}(y),
\end{equation}
each direction satisfies a one-dimensional Ermakov equation,

\begin{equation}
	X''+k_x^2X
	=
	\frac{C_{x0}^2}{\hbar^2X^3},
	\qquad
	Y''+k_y^2Y
	=
	\frac{C_{y0}^2}{\hbar^2Y^3},
\end{equation}
with
\[
k_x^2+k_y^2
=
\frac{2mE}{\hbar^2}.
\]
The stationary amplitudes are constructed in the
Ermakov form,
\begin{equation}
	X^2(x)=A_{k_x}\left(1+\varepsilon_{k_x}\sin^2(k_x x)\right),
	\qquad
	Y^2(y)=A_{k_y}\left(1+\varepsilon_{k_y}\sin^2(k_y y)\right),
\end{equation}
giving
\begin{equation}
	R^2(x,y)=A_{k_x}A_{k_y}
	\left(1+\varepsilon_{k_x}\sin^2(k_x x)\right)
	\left(1+\varepsilon_{k_y}\sin^2(k_y y)\right).
\end{equation}
The phase separates as
\begin{equation}
	S_{k_x}=\hbar k_x x+\frac{\hbar\varepsilon_{k_x}}{4}\sin(2k_x x),
	\qquad
	S_{k_y}=\hbar k_y y+\frac{\hbar\varepsilon_{k_y}}{4}\sin(2k_y y).
\end{equation}

\noindent Using the one-dimensional Fourier--Bessel coefficients
\(\mathcal C_u(\varepsilon_{k_x})\) and
\(\mathcal C_v(\varepsilon_{k_y})\),
the separable rectangular aperture wavefunction admits
the spectral form
\begin{equation}
	\psi(x,y)
	=
	\sqrt{A_{k_x}A_{k_y}}
	\sum_{u,v}
	\mathcal C_u(\varepsilon_{k_x})
	\mathcal C_v(\varepsilon_{k_y})
	e^{i(2u+1)k_xx}
	e^{i(2v+1)k_yy}.
\end{equation}

\noindent For a rectangular aperture centred at the origin with
characteristic width \(L\), Dirichlet-type boundary
conditions quantise the admissible wave numbers according to
\[
k_x
=
\frac{u\pi}{L},
\qquad
k_y
=
\frac{v\pi}{L},
\qquad
u,v=1,2,\dots,
\]
so that the stationary energy satisfies
\[
\frac{\pi^2}{L^2}
(u^2+v^2)
=
\frac{2mE}{\hbar^2}.
\]

\noindent The rectangular construction therefore provides the
natural separable starting point for examining translated
and reduced aperture geometries, which are developed in
the following section.
\vspace{0.5pc}

\section{Rectangular to parabolic slit reduction}
\label{sec:Sec5}
We now consider a reduced aperture geometry obtained
from the separable rectangular construction developed
in Section~\ref{sec:Sec4}. In the narrow-aperture
regime, one transverse direction dominates the
propagation, allowing the two-dimensional Ermakov
structure to be reduced to an effective one-dimensional
description. This reduction naturally gives rise to a
parabolic slit geometry with Fresnel--type phase
modulation and spectrally organised sideband structure.

\subsection{Parabolic slit reduction and optical superposition}
\label{sec:parabolicred}
\vspace{0.5pc}

\noindent For a narrow rectangular aperture, we exploit the separation of scales between the transverse directions. In the $x$-direction, the slit width is sufficiently small that amplitude variations across the aperture are negligible, and we set
\[
R_{k_x}(x) \simeq A_{k_x}, \qquad \varepsilon_{k_x} \to 0.
\]
The dominant spatial modulation is therefore carried along the $y$-direction. We consider a slit centred along the \(y\)-axis, with propagation into the region \(x>0\). Near the optical axis, the wavefront is approximated locally by a parabolic reduction.
\vspace{0.5pc}

\noindent We impose a parabolic reduction near the central axis,
\begin{equation}
	x^2 + y^2 = \mathcal{R}^2,
	\qquad
	y(x) \simeq \mathcal{R} - \frac{x^2}{2\mathcal{R}},
	\qquad |x| \ll \mathcal{R}.
\end{equation}
Under this reduction, the effective amplitude becomes
\begin{equation}
	R_{\rm eff}^2(x)
	=
	A_{k_x}
	\left[
	1 + \varepsilon_{k_x}
	\sin^2\!\left(
	k_y \mathcal{R} - \frac{k_y x^2}{2\mathcal{R}}
	\right)
	\right].
\end{equation}
This may be written as
\begin{equation}
	R_{\rm eff}^2(x)
	=
	A_{k_x}
	\left[
	1 + \frac{\varepsilon_{k_x}}{2}
	-
	\frac{\varepsilon_{k_x}}{2}
	\cos\!\left(
	2k_y \mathcal{R} - \frac{k_y x^2}{\mathcal{R}}
	\right)
	\right].
\end{equation}
\noindent The effective phase may be written in closed form as
\begin{equation}
	S_{\rm eff}(x)
	=
	\hbar k_x x
	+
	\hbar k_y \mathcal{R}
	-
	\frac{\hbar k_y}{2\mathcal{R}} x^2
	+
	\frac{\hbar \varepsilon_{k_x}}{4}
	\sin\!\left(
	2k_y \mathcal{R} - \frac{k_y x^2}{\mathcal{R}}
	\right).
\end{equation}
Applying the Jacobi-Anger expansion to the nonlinear phase term gives
\begin{equation}
	e^{iS_{\rm eff}/\hbar}
	=
	e^{ik_xx}
	e^{ik_y\mathcal{R}}
	e^{-ik_yx^2/(2\mathcal{R})}
	\sum_{n=-\infty}^{\infty}
	J_n\!\left(\frac{\varepsilon_{k_x}}{4}\right)
	\exp\!\left[
	in\left(
	2k_y\mathcal{R}-\frac{k_yx^2}{\mathcal{R}}
	\right)
	\right].
\end{equation}
where the factors respectively encode the carrier, Fresnel, and nonlinear modulation contributions.
\vspace{0.5pc}

\noindent The term \(\hbar k_x x\) acts as a slowly varying
transverse carrier phase across the slit and contributes
only weakly to the dominant propagation dynamics.
\vspace{0.5pc}

\noindent
Therefore the reduced single-slit wavefunction takes the amplitude--phase form
\begin{equation}
	\begin{aligned}
		\psi_{\rm eff}(x)
		&=
		\sqrt{
			A_{k_x}\left[
			1+\frac{\varepsilon_{k_x}}{2}
			-\frac{\varepsilon_{k_x}}{2}
			\cos\!\left(
			2k_y\mathcal{R}-\frac{k_yx^2}{\mathcal{R}}
			\right)
			\right]
		}
		\,e^{ik_xx}e^{ik_y\mathcal{R}}e^{-ik_yx^2/(2\mathcal{R})}
		\\
		&\quad\times
		\sum_{n=-\infty}^{\infty}
		J_n\!\left(\frac{\varepsilon_{k_x}}{4}\right)
		\exp\!\left[
		in\left(
		2k_y\mathcal{R}-\frac{k_yx^2}{\mathcal{R}}
		\right)
		\right].
	\end{aligned}
\end{equation}

\noindent For weak modulation,
\(
|\varepsilon_{k_x}|\ll1
\),
the amplitude term under the square root may itself be
expanded spectrally, so that each nonlinear sideband
inherits an effective coefficient
\(\mathcal C_n\)
together with a quadratic Fresnel-type chirp factor. The corresponding \(n\)-th sideband therefore carries
an effective quadratic chirp phase of the form
\[
\exp\!\left[
-i
\left(
n+\frac12
\right)
\frac{k_yx^2}{\mathcal R}
\right],
\]
showing that each spectral component inherits a
distinct Fresnel-type phase modulation.
\vspace{0.5pc}

\noindent Consequently, the reduced slit geometry admits a
spectrally organised nonlinear diffraction structure,
in which each sideband carries a distinct chirped phase.
Thus the parabolic reduction retains a residual geometric nonlinearity: the two-dimensional spatial spread is encoded through the square-root amplitude envelope and the Fresnel-type nonlinear phase chirp.

\subsection{Spectral superposition of two shifted sources in one dimension}
\vspace{0.5pc}

We now return to the one-dimensional parent branch of Section~\ref{sec:Sec3} and consider two spatially shifted copies. Independent of the parabolic slit reduction, this construction reveals the translation-covariant structure of the Fourier--Bessel representation.
\begin{equation}
	\psi_1(x)=\psi\!\left(x+\frac{a}{2}\right),
	\qquad
	\psi_2(x)=\psi\!\left(x-\frac{a}{2}\right).
\end{equation}
Using the canonical form of the one-dimensional representation given by ~\eqref{eq:jaexp1d}, we get
\begin{equation}
	\psi\!\left(x\pm\frac{a}{2}\right)
	=
	\sqrt{A}
	\sum_{n=-\infty}^{\infty}
	\mathcal{C}_n(\varepsilon)
	e^{i(2n+1)kx}
	e^{\pm i(2n+1)ka/2}.
\end{equation}
The symmetric and antisymmetric combinations are 
\begin{equation}
	\Psi(x):=\psi_1+\psi_2 = 2\sqrt{A}
	\sum_{n=-\infty}^{\infty}
	\mathcal{C}_n(\varepsilon)
	\cos\!\left[(2n+1)\frac{ka}{2}\right]
	e^{i(2n+1)kx},
\end{equation}
and
\begin{equation}
	\mathcal{X}(x):=\psi_1-\psi_2 = 2i\sqrt{A}
	\sum_{n=-\infty}^{\infty}
	\mathcal{C}_n(\varepsilon)
	\sin\!\left[(2n+1)\frac{ka}{2}\right]
	e^{i(2n+1)kx}.
\end{equation}

\noindent The symmetric and antisymmetric branches isolate the even and odd spectral interference structures respectively. This decomposition separates constructive and destructive interference within the Fourier--Bessel representation.
\vspace{0.5pc}

\noindent This representation makes the shifted-source structure explicit: the Ermakov modulation determines the Fourier--Bessel coefficients ${\mathcal C_n(\varepsilon)}$, while the source separation $a/2$ weights each spectral harmonic through the interference factors
\begin{equation}
	\cos\left[\frac{(2n+1)ka}{2}\right],
	\quad
	\sin\left[\frac{(2n+1)ka}{2}\right],
\end{equation}
while preserving the underlying Fourier--Bessel harmonic structure.
\vspace{0.5pc}

\noindent The resulting Fourier--Bessel harmonic structure also suggests a natural extension to two-dimensional slit geometries through the parabolic reduction developed in Section~\ref{sec:parabolicred}. In this setting, interference between shifted sources inherits the corresponding Fresnel--type chirp, which is captured through the symmetric and antisymmetric spectral branches.
\vspace{0.5pc}

\noindent From a wave-propagation perspective, the shifted-source construction therefore provides a simplified two-slit analogue in which interference arises through spectrally organised amplitude and phase modulation. The resulting symmetric and antisymmetric branches encode the chirped interference structure expected from near-field diffraction and nonlinear optical systems~\cite{BornWolf}, while remaining analytically tractable within the stationary Bohmian framework.

\section{Summary}
\label{sec:Sec6}
\vspace{0.5pc}

\noindent
The preceding analysis shows that stationary
superposition in the Bohmian description undergoes a
spectral reorganisation arising from the underlying
nonlinear amplitude--phase dynamics. Under weakly
modulated stationary conditions, the mean branch
preserves an Ermakov--Pinney backbone governed by a
Wronskian invariant, while the difference branch
acquires a Mathieu--Hill type modulation.
\vspace{0.5pc}

\noindent
Despite the nonlinear amplitude--phase coupling, the
stationary branches admit a linearly reconstructible
Fourier--Bessel spectral representation. From a
quantum-mechanical perspective, this provides a
correspondence between linear wavefunction addition in
the Schr\"odinger picture and nonlinear amplitude--phase
organisation in the Bohmian picture. The resulting
spectral structure includes modulation-driven
sidebands, quadratic phase chirp and
translation-dependent interference effects, while
remaining consistent with the overall spectral
organisation~\cite{Boyd,Gilberto}.

\section{Conclusions}
\label{sec:Sec7}
\vspace{0.5pc}

In this work, we have examined the problem of
superposition in Bohmian mechanics, where amplitude
and phase obey coupled nonlinear equations. Unlike
the linear Schr\"{o}dinger framework, superposition is
not naturally closed at the level of these variables,
but instead emerges under controlled stationary
conditions through a spectral reconstruction.
\vspace{0.5pc}

\noindent The main results can be summarised as follows:
\vspace{0.5pc}

\noindent (i) Nonlinear superposition structure.
Under controlled near-degenerate stationary conditions, the system admits a hierarchical decomposition. The mean amplitude evolves according to a closed Ermakov--Pinney equation governed by a Wronskian invariant, while the difference amplitude obeys a parametrically driven Mathieu--Hill type equation. Even for weak perturbations of conventional stationary superposition, nonlinearity persists through the modulated amplitude difference and its coupled phase dynamics.
\vspace{0.5pc}

\noindent (ii) Emergent linearity.
Although the underlying equations are nonlinear, a linear superposition structure re-emerges through a spectral representation. By reconstructing a weakly modulated stationary branch, the phase-modulated state admits a Jacobi--Anger expansion, yielding a Fourier--Bessel decomposition with square-summable coefficients.
\vspace{0.5pc}

\noindent	(iii) Spectral–spatial equivalence.
The Bessel coefficients define a discrete momentum distribution, while their norm matches the spatially averaged amplitude envelope. This establishes a correspondence between nonlinear amplitude modulation and linear spectral organisation.
\vspace{0.5pc}

\noindent(iv) Translation covariance. In the one-dimensional shifted-source construction, spatial translations act as phase rotations in the Fourier--Bessel representation, leading to symmetric and antisymmetric spectral weights while preserving the underlying orthogonal harmonic structure.
\vspace{0.5pc}

\noindent (v) Wave-geometric interpretation.
The framework applies naturally to bounded systems such as rectangular apertures, where separability leads to a Fourier--Bessel lattice structure. Through parabolic reduction and shifted-source constructions, the resulting amplitude--phase dynamics exhibits Fresnel--type phase chirp and modulation-driven sidebands, providing a simplified two-slit analogue within the stationary Bohmian setting.
\vspace{0.5pc}

\noindent
Overall, the present construction shows that a linear spectral superposition structure re-emerges from the underlying nonlinear amplitude--phase dynamics of stationary Bohmian systems. Through Wronskian-governed phase organisation, controlled modulation and spectral reconstruction, the framework establishes an analytical bridge between nonlinear Bohmian dynamics and the linear superposition structure of conventional quantum mechanics.
\vspace{0.5pc}

\noindent\textbf{Funding Statement.}
The author received no external funding for this work.

\vspace{0.5pc}

\noindent\textbf{Data Availability Statement.}
No datasets were generated or analysed during the current study.

\vspace{0.5pc}

\noindent\textbf{Ethics Statement.}
This work does not involve human participants, animals, or identifiable personal data.

\vspace{0.5pc}

\noindent\textbf{Declaration of Interests.}
The research is an independent work of the author and declares no conflicts of interest with the employer, IBM Research, Albany, NY, USA.

\vspace{0.5pc}

\noindent\textbf{Use of Generative AI.}
Generative AI tools were used in the preparation of this manuscript solely for language editing, grammar refinement, and equation cross-verification. These tools were not used to generate the scientific results, derivations, or core interpretations presented here. All mathematical checks, technical conclusions, and final editorial decisions remain the sole responsibility of the author.

\appendix
\renewcommand{\thesection}{Appendix~\Alph{section}}
\renewcommand{\theequation}{\Alph{section}\arabic{subsection}.\arabic{equation}}
\setcounter{equation}{0}

\section{Fourier--Bessel representation}
\label{app:fourier-bessel}
\vspace{0.5pc}

The parent stationary branch is represented in polar form as
\begin{equation}
	\psi(x)=R(x)\exp\!\left(\frac{i}{\hbar}S(x)\right),
\end{equation}
where, to first order in \(\varepsilon\),
\begin{equation}
	R(x)\simeq
	\sqrt{A}\left[
	1+\frac{\varepsilon}{2}\sin^2(kx)
	\right],
\end{equation}
and
\begin{equation}
	\exp\!\left(\frac{i}{\hbar}S(x)\right)
	=
	\sum_{n=-\infty}^{\infty}
	J_n\!\left(\frac{\varepsilon}{4}\right)
	e^{i(2n+1)kx}.
\end{equation}
Using
\begin{equation}
	1+\frac{\varepsilon}{2}\sin^2(kx)
	=
	1+\frac{\varepsilon}{4}
	-\frac{\varepsilon}{8}e^{2ikx}
	-\frac{\varepsilon}{8}e^{-2ikx},
\end{equation}
the stationary branch therefore admits the compact
Fourier--Bessel representation
\begin{equation}
	\psi(x)=
	\sqrt{A}
	\sum_{n=-\infty}^{\infty}
	\mathcal C_n(\varepsilon)\,
	e^{i(2n+1)kx},
\end{equation}
where
\begin{equation}
	\boxed{	\mathcal C_n(\varepsilon)
		=
		\left(1+\frac{\varepsilon}{4}\right)
		J_n\!\left(\frac{\varepsilon}{4}\right)
		-\frac{\varepsilon}{8}
		\left[
		J_{n-1}\!\left(\frac{\varepsilon}{4}\right)
		+
		J_{n+1}\!\left(\frac{\varepsilon}{4}\right)
		\right].
	}		
\end{equation}

\subsection{Fourier--Bessel coefficient properties}
\vspace{0.5pc}

The modulated amplitude--phase structure leads to the
Fourier--Bessel coefficients
\begin{equation}
	\mathcal C_n(\varepsilon)
	=
	\left(1+\frac{\varepsilon}{4}\right)
	J_n\!\left(\frac{\varepsilon}{4}\right)
	-\frac{\varepsilon}{8}
	\left[
	J_{n-1}\!\left(\frac{\varepsilon}{4}\right)
	+
	J_{n+1}\!\left(\frac{\varepsilon}{4}\right)
	\right].
\end{equation}
Using the Bessel recurrence relation
\begin{equation}
	J_{n-1}(z)+J_{n+1}(z)=\frac{2n}{z}J_n(z),
	\qquad z=\frac{\varepsilon}{4},
\end{equation}
this expression simplifies to the principal-index form
\begin{equation}
	\mathcal C_n(\varepsilon)
	=
	\left(1+\frac{\varepsilon}{4}-n\right)
	J_n\!\left(\frac{\varepsilon}{4}\right).
\end{equation}
\vspace{0.5em}
\subsection{Norm and convergence of the Fourier--Bessel coefficients}
\vspace{0.5em}
The squared coefficient becomes
\begin{equation}
	\label{eq:fourierbesselsingle}
	|\mathcal C_n(\varepsilon)|^2
	=
	\left(1+\frac{\varepsilon}{4}-n\right)^2
	J_n^2\!\left(\frac{\varepsilon}{4}\right).
\end{equation}
Hence the coefficient norm is
\begin{equation}
	\sum_{n=-\infty}^{\infty}
	|\mathcal C_n(\varepsilon)|^2
	=
	\sum_{n=-\infty}^{\infty}
	\left(1+\frac{\varepsilon}{4}-n\right)^2
	J_n^2\!\left(\frac{\varepsilon}{4}\right).
\end{equation}
Using the standard identities
\begin{equation}
	\sum_{n=-\infty}^{\infty}J_n^2(z)=1,
	\qquad
	\sum_{n=-\infty}^{\infty}nJ_n^2(z)=0,
	\qquad
	\sum_{n=-\infty}^{\infty}n^2J_n^2(z)=\frac{z^2}{2},
\end{equation}
we obtain
\begin{equation}
	\sum_{n=-\infty}^{\infty}
	|\mathcal C_n(\varepsilon)|^2
	=
	1+\frac{\varepsilon}{2}
	+\frac{3\varepsilon^2}{32}.
\end{equation}
\vspace{0.5em}
\subsection{Consistency of the modulated amplitude representation}
\vspace{0.5em}
To first order in $\varepsilon$,
\begin{equation}
	\sum_{n=-\infty}^{\infty}
	|\mathcal C_n(\varepsilon)|^2
	\simeq
	1+\frac{\varepsilon}{2}.
\end{equation}
This matches the spatial average of the amplitude envelope,
\begin{equation}
	R^2(x)=A\left(1+\varepsilon\sin^2(kx)\right),
\end{equation}
since
\begin{equation}
	\left\langle
	1+\varepsilon\sin^2(kx)
	\right\rangle
	=
	1+\frac{\varepsilon}{2}.
\end{equation}
The series is absolutely convergent because, for fixed \(z\),
\begin{equation}
	J_n(z)\sim \frac{1}{n!}\left(\frac{z}{2}\right)^n,
	\qquad n\to \infty,
\end{equation}
so the polynomial prefactor
\[
\left(
1+\frac{\varepsilon}{4}-n
\right)^2
\]
grows only algebraically and is therefore dominated by the
factorial decay of \(J_n^2(z)\). Therefore
\begin{equation}
		\sum_{n=-\infty}^{\infty}
		|\mathcal C_n(\varepsilon)|^2 < \infty.
\end{equation}

\noindent 
The Fourier--Bessel representation therefore defines
a square-summable coefficient sequence and admits
a consistent quantum-mechanical normalisation.
\vspace{0.5pc}

\end{document}